\documentclass[aps,prl,showpacs,amsmath,amssymb,superscriptaddress,floatfix,twocolumn]{revtex4-1}
\usepackage{graphicx}
\usepackage{epstopdf}
\usepackage{amsmath,amssymb}
\usepackage{color}
\usepackage{CJK}

\newcommand{\wpcmsq}{W/cm$^2$\,}
\newcommand{\OO}{O$_2$\,}
\newcommand{\COO}{CO$_2$\,}
\newcommand{\COOp}{CO$_2^{+}$\,}
\newcommand{\COOpp}{CO$_2^{2+}$\,}
\newcommand{\COp}{CO$^+$\,}
\newcommand{\Op}{O$^+$\,}
\newcommand{\OOp}{O$_2^+$\,}
\newcommand{\Cp}{C$^+$\,}

\begin{document}

\title{Direct Observation of Molecular Oxygen Production from Carbon Dioxide}

\author{Seyedreza Larimian}
\author{Sonia Erattupuzha}
\affiliation{Photonics Institute, Vienna University of Technology, A-1040 Vienna, Austria}
\author{Sebastian Mai}
\author{Philipp Marquetand}
\author{Leticia Gonz\'alez}
\affiliation{Institute of Theoretical Chemistry, University of Vienna, A-1090 Vienna, Austria}
\author{Andrius Baltu\v{s}ka}
\author{Markus Kitzler}
\author{Xinhua Xie}
\email[Electronic address: ]{xinhua.xie@tuwien.ac.at}
\affiliation{Photonics Institute, Vienna University of Technology, A-1040 Vienna, Austria}

\date{\today}

\begin{abstract}
Oxygen (\OO) is one of the most important elements required to sustain life.
The concentration of \OO on Earth has been accumulated over millions of years and has a direct connection with that of \COO.
Further, \COO plays an important role in many other planetary atmospheres.
Therefore, molecular reactions involving \COO are critical for studying the atmospheres of such planets.
Existing studies on the dissociation of \COO are exclusively focused on the C--O bond breakage.
Here we report first experiments on the direct observation of molecular Oxygen formation from \COO in strong laser fields with a reaction microscope.
Our accompanying simulations suggest that \COO molecules may undergo bending motion during and after strong-field ionization which supports the molecular Oxygen formation process.
The observation of the  molecular Oxygen formation from \COO may trigger further experimental and theoretical studies on such processes with laser pulses, and provide hints in studies of the \OO and \OOp abundance in \COO-dominated planetary atmospheres.
\end{abstract}

\maketitle


\OO production is one of the most important processes for the biosphere of the Earth.
Oxygen molecules are mainly generated via the photosynthesis by green plants and algae from carbon dioxide and water:\textit{n}\COO+\textit{n}H$_2$O$\xrightarrow[]{light}$(CH$_2$O)$_\textit{n}$+\textit{n}\OO ~\cite{singhal2012concepts}. 
\COO is not only important for the atmosphere on Earth, it is also the dominant compound of the atmosphere on other planets, such as Mars and Venus.
One of the most crucial tasks for the quest to establish a human settlement on Mars is the production of \OO \cite{hecht2015mars}.
Because more than 95\% of the atmosphere on Mars is \COO, it will be extremely helpful if \OO can be produced directly from \COO.
In the past, it was observed that dissociation of \COO via absorption of photons leads to carbon monoxide (CO) and oxygen atoms (O)~\cite{chueh2010high}.
However, theoretical simulations suggested the possibility of generating \OO through the dissociation of a \COO molecule \cite{Hwang2000,*Grebenshchikov2013}.
A recent experiment showed the evidence of \OO formation from \COO molecules after UV excitation through the detection of C$^+$~\cite{Lu2014}.
So far, \OO formation from \COO has not been directly observed.

In the past decade, intense ultrashort laser pulses have been successfully applied to trigger and control molecular reactions such as dissociation and isomerization \cite{Itakura2003,kling2006,znak09prl,xie12_2,Dominik2012,Alnaser2014,xie2014prl,xie2014prx,Marquetand2014,Miura2014,xie2015sr}.
When a molecule interacts with a strong laser field, electrons from outer molecular orbitals can be excited or removed through tunneling or over-the-barrier ionization which may prepare the molecule in an excited state or a state with a certain charge.
As a consequence, the excited or ionized molecule may undergo severe geometrical reconfiguration and may also break into several fragments or form new chemical bonds.
Because of the importance of \COO in many research disciplines, strong-field induced reactions of \COO have been experimentally studied with ultrashort lasers by several research groups. However, these studies mainly focused on the topic of ionization and dissociation \cite{Bryan2000,*Brichta2007,*Bocharova2011,*wu2012,*Erattupuzha2016}.
In this paper, for the first time to our knowledge, we report on the direct observation of \OOp formation from \COO induced by strong laser pulses with a reaction microscope.
Previous studies revealed that neutral \OO molecules can be conveniently obtained through the neutralization of \OOp at metal surfaces \cite{Reijnen1988,*Lorente1997}, which makes the \OOp formation method presented here a candidate for producing breathable \OO direct from \COO.
Although the efficiency of the reaction leading to \OOp formation is rather low, our results can serve as a proof-of-principle experiment on the laser-induced \OOp formation directly from \COO.
Our quantum chemical simulations indicate that the bending motion during and after the strong field interaction plays a critical role in the \OOp formation process.

In the experiments we performed coincidence measurements of ions from doubly ionized \COO with a reaction microscope \cite{ullrich03,*doerner00}.
Laser pulses with a pulse duration (full width at half maximum of the peak intensity) of 25 fs, a central wavelength of 790 nm, a repetition rate of 5 kHz and peak laser intensities on the order of $10^{14}$ \wpcmsq are produced with a home-built femtosecond Ti:Sa laser amplifier system.
The reaction microscope consists of a two-stage gas jet arrangement to provide an internally cold ultrasonic gas jet of \COO with a diameter of about 170 $\mu$m, and an ultra-high vacuum interaction chamber ($1.\times10^{-10}$ mbar).
The laser beam is focused in the interaction chamber onto the gas jet with a spherical silver mirror with a focal length of 60 mm.
A homogenous DC field of 10.55 V/cm is applied along the axis of the TOF spectrometer to accelerate positively charged particles to a position-sensitive detector.
The laser field is linearly polarized along the spectrometer axis.
With the measured position and TOF data, the three-dimensional momentum vector of a certain ion can be retrieved.
All possible two-body fragmentation processes can be distinguished using the momentum conservation selection to a pair of ionic fragments.
The laser intensities of our measurements are calibrated using the shape of the TOF spectrum of protons from the dissociation of H$_2^+$ in the strong laser field~\cite{Alnaser_2004}.
More details of the experimental setup can be found in our previous publications \cite{Zhang2012,xie12prl}.

\begin{figure}[htbp]
\centering
\includegraphics[width=0.4\textwidth,angle=0]{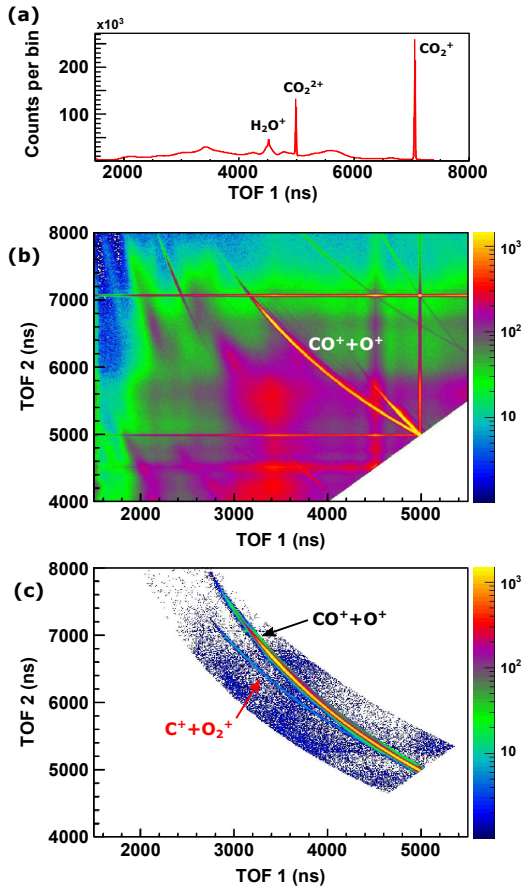}
\caption{ (a) Measured time-of-flight spectrum of ions. (b,c) Measured photo-ion-photo-ion coincidence distribution without (b) and with (c) coincidence selection.} \label{fig:pipico}
\end{figure}
We start by focusing on the identification of \OOp formation from \COOpp in our measurements.
In Fig.~\ref{fig:pipico}(a) we present the time-of-flight (TOF) spectrum of ions measured when 25 fs laser pulses with a peak intensity of 1.6$\times 10^{14}$ \wpcmsq interact with isolated \COO molecules. The peaks of singly and doubly ionized \COO molecules are clearly visible.
Two-body fragmentation channels can be unambiguously identified in the photo-ion-photo-ion coincidence (PIPICO) distribution.
Due to momentum conservation, two particles from the same molecule exhibit a correlation between their TOFs, which leads to sharp parabolic lines in the PIPICO distribution, as shown in Fig.~\ref{fig:pipico}(b).
In this figure, the strongest parabolic line is identified as the \COp+\Op fragmentation channel.
The parabolic line corresponding to a weak fragmentation channel can be enhanced by applying coincidence conditions using the retrieved momentum vectors.
The PIPICO distribution with coincidence selection perpendicular to the spectrometer axis is depicted in Fig.~\ref{fig:pipico}(c), in which the signal corresponding to the formation of oxygen molecule appears with a high signal-to-noise ratio.
Our measurements show that the yield of the \OOp channel is about three orders of magnitude weaker than that of the \COp channel.
Nevertheless, we have directly observed the production of oxygen molecule from doubly ionized \COO in a strong laser field.

\begin{figure}[ht]
\centering
\includegraphics[width=0.48\textwidth,angle=0]{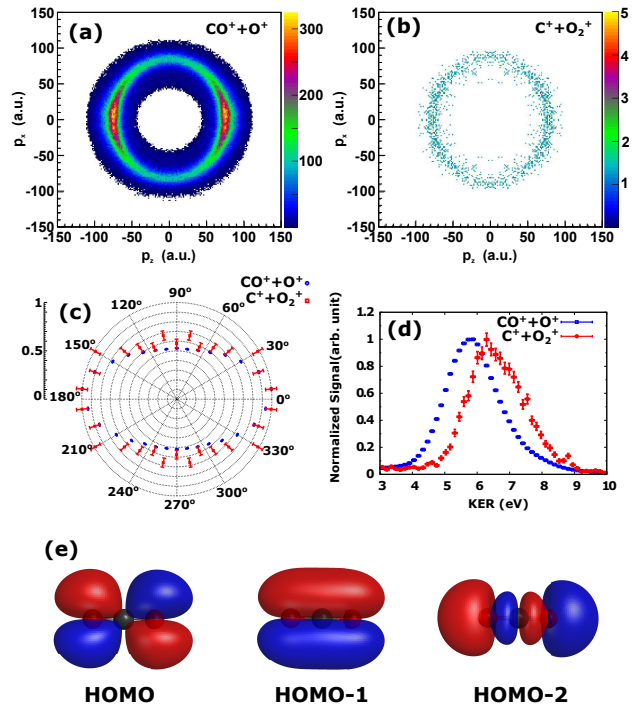}
\caption{Cut through the measured momentum distributions along the laser pulse propagation direction (x) and the polarization axis (z) for the \COp channel (a) and the \OOp channel (b) with $|p_y-30|<5$ a.u.. (c) Measured angular distribution derived from (a) and (b), respectively, for the \COp channel (blue points) and the \OOp channel (red points). (d) Kinetic energy release distribution for the \COp channel (blue points) and the \OOp channel (red points). (e) Molecular orbitals of \COO calculated with the quantum chemical software {\sc gamess} \cite{gamess}.} \label{fig:ker}
\end{figure}
From the data selected with the coincidence condition we retrieved the three-dimensional momentum vectors of the two involved ions and the kinetic energy release (KER) of the two observed fragmentation channels.
The measured momentum distributions of the \COp and the \OOp in the laser polarization plane are presented in Figs.~\ref{fig:ker}(a) and (b), respectively.
It can be seen that for both channels the signals are mainly distributed along the laser polarization direction.
To see this more clearly, we plot the angular distributions obtained by integrating the measured signals over radial and azimuthal coordinates in Fig.~\ref{fig:ker}(c).
The angular distribution contains information on the molecular orbitals involved in the ionization process \cite{xie2014prx}.
The calculated three highest occupied molecular orbitals (HOMOs) are illustrated in Fig.~\ref{fig:ker}(e) \cite{gamess}.
According to the Molecular-Ammosov-Delone- Krainov (MO-ADK) theory \cite{tong2002theory}, the probabilities for ionization from HOMO and HOMO-1 peak along the direction perpendicular to the molecular axis, whereas that from HOMO-2 peaks along the molecular axis.
The measured angular distributions [Fig.~\ref{fig:ker}(c)] indicate that the removal of electrons from HOMO-2 might be involved in the double ionization of \COO before both fragmentation processes.

Further information about the involved states before the fragmentation can be obtained from the KER distribution of the two-body fragmentation channels, presented in Fig.~\ref{fig:ker}(d).
The KER value is given by the energy difference between the initially created dication and the final fragmentation products.
The mean values of the KER are 5.9 eV and 6.6 eV for the \COp channel and the \OOp channel, respectively.
The width of the KER distribution is related to both the width of the nuclear wave packet and the potential energy surfaces involved in the ionization process.
The measured FWHMs are about 2.0 eV for the both channels, which may suggest that both fragmentation channels originate from an electronic state with a similar curvature.

\begin{figure}[ht]
\centering
\includegraphics[width=0.4\textwidth,angle=0]{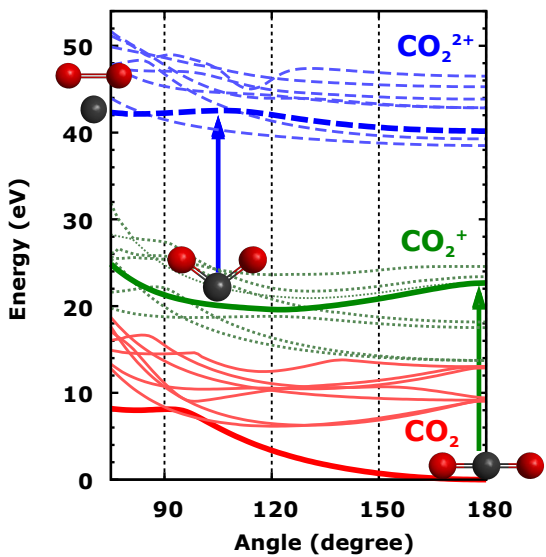}
\caption{Simulated potential energy curves for the \COO neutral (solid red lines), the cation (dotted green lines) and the dication (dashed blue lines) over the bending angle. A possible pathway leading to \OOp formation is indicated with thick potential energy curves and arrows.} \label{fig:pecs}
\end{figure}
To gain insight into the \OOp formation mechanism from \COOpp, we performed quantum chemical simulations.
In order to obtain a balanced and accurate description of all electronic states, we employed CASSCF (complete active space self-consistent field), augmented by MS-CASPT2 (multi-state complete active space perturbation theory second order) to include electronic correlation effects ~\cite{Pulay2011IJQC,*Andersson1995,*Roos2005,*Finley1998CPL}.
The calculations of potential energy curves were performed within $C_{2v}$ symmetry, with the twofold rotation axis perpendicular to the line connecting the two oxygen atoms.
In this setup, the active space contained 4 orbitals of $a_1$ symmetry, 3 orbitals of $b_1$, 1 orbital of $a_2$ and 2 orbitals of $b_2$ symmetry.
In the active space, all possible distributions of 12, 11 or 10 electrons were considered, respectively, while 10 electrons occupied the $2s$ orbitals of oxygen and all $1s$ orbitals.
The calculations were performed with \textsc{Molcas} 8.0 \cite{Aquilante2015JCC} and the ANO-RCC-VTZP basis set.
Note that the calculations did only consider the 2 lowest states in each symmetry, while Rydberg states were not considered.
Figure~\ref{fig:pecs} shows the potential energy curves of these 24 states (8 neutral, 8 cationic, 8 dicationic) over the bending angle.

To form \OO from \COO the molecule should first form a triangular geometry.
However, as shown in Fig.~\ref{fig:pecs}, the low-energy dicationic states (i.e., below approximately 46~eV) exhibit minima at the linear geometry, thus providing no force to form a bent geometry.
Hence, it is unlikely that the triangular geometry is formed in the dicationic states after the laser field has faded.
More likely, therefore, is the formation of the triangular species during the strong field interaction.
There are a number of neutral and cationic states with a gradient pointing towards a bent geometry.
Nuclear wave packet dynamics in these excited, intermediate states \cite{Sandor2016PRL} could thus lead to the bending of the molecule, eventually resulting in the formation of a triangular intermediate.

In Fig.~\ref{fig:pecs}, we highlight a possible excitation pathway for the dynamics leading to \OOp formation:
the \COO is initially in the neutral ground state (thick red line).
Through strong field nonresonant ionization the system is promoted to an excited state of \COOp (thick green line), which has a minimum at an angle of approximately 120$^\circ$.
Hence, after the first ionization step the nuclear wave packet evolves towards smaller bending angles, and the molecule starts to bend.
When the wave packet approaches the mentioned minimum, the molecule is further ionized to an excited state of \COOpp (thick blue line), which provides a driving force towards even smaller bond angles.
Eventually, in the dicationic state the molecule forms a triangular intermediate, which may release \Cp to form \OOp.

In order to investigate whether bending in \COO could be fast enough to occur within the laser pulse duration, we performed simulations on field-free semi-classical dynamics in the first neutral excited state which has a similar shape as the highlighted state of \COOp.
Indeed, within 25~fs a bending angle of about 110$^\circ$ was reached, which is consistent with our hypothesis that a triangular geometry can form within the pulse duration.

One should note, however, that due to the influence of the strong laser field, the relevance of the potential energy curves shown in Figure~\ref{fig:pecs} is somewhat limited.
In particular, due to the Stark effect the potentials may be significantly distorted in a complicated, time-dependent manner.
The Stark effect and the difficulties of accurately describing resonant multi-photon absorption and ionization make any prediction of the dynamics of \COO during the presence of the laser field extremely difficult.
Hence, the potential energy curves only serve a qualitative understanding of the \OOp formation from the \COO dication.

\begin{figure}[ht]
\centering
\includegraphics[width=0.4\textwidth,angle=0]{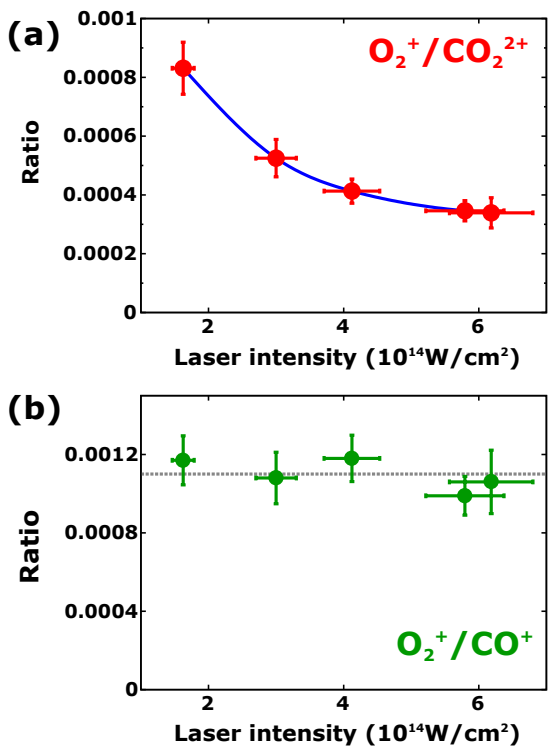}
\caption{ (a) Ratio between the yields of \OOp and that of stable \COOpp as a function of the laser peak intensity. (b) Ratio between the yields of \OOp and \COp channels as a function of the laser peak intensity. The blue line in (a) and the gray line in (b) are used to guide the eyes. } \label{fig:int}
\end{figure}
A interesting aspect of the formation of \OOp from \COO is the efficiency of this process compared to the \COp channel.
Previously, it has been demonstrated that molecular reaction dynamics such as bond breakage and bond formation can be controlled by the parameters of the applied laser pulse \cite{Shapiro2003,Itakura2003,Daniel2003,Marquetand2007,*Marquetand2008,xie12_2,xie2014prl,xie2014prx,Alnaser2014,Miura2014,*xie2015sr}.
One of the most important parameters is the peak of a laser pulse.
It has been shown that the peak intensity can be used to control double ionization pathways \cite{Zhang2012,xie2014prx}.

In the present measurements, we adjusted the laser intensity by reflecting off a certain amount of the laser beam using one or more pellicles.
Fig.~\ref{fig:int} illustrates the efficiency of \OOp production in comparison with that of \COOpp and \COp.
An obvious intensity dependence of the \OOp yield relative to that of \COOpp is visible in Fig.~\ref{fig:int}(a).
With the increase of the laser peak intensity from 1.7 to 6.1 $\times 10^{14}$\wpcmsq, the relative \OOp yield decreases by a factor of about 2.5.
In contrast, the yield of the \OOp normalized to that of the \COp channel has no apparent dependence on the laser intensity [Fig.~\ref{fig:int}(b)].
As already shown in previous sections, the \OOp channel and the \COp channel have similar angular dependence and exhibit the same width of the KER distributions.
Together with independence of the relative channel strength on the laser intensity these observations imply that the dynamics leading to the two channels may happen on the same potential energy surfaces.
A better understanding of the origin of the \OOp formation may be obtained from pump-probe experiments with few-cycle laser or XUV attosecond pulses. Such experiments should be able to temporally resolve the process of \OOp formation from \COOpp.

In conclusion, for the first time in an experiment we observed the production of \OOp from \COO driven by a strong laser pulse with a reaction microscope. Our accompanying quantum chemical simulations suggest that the bending motion during the strong-field interaction may trigger the \OOp formation.
The similarity of the measured angular distributions, the width of the KER distribution and the dependence of the relative yields on the laser intensity indicate that the process of \OOp formation may originate from the same intermediate species prepared by the strong field ionization as the dissociation to \COp and \Op.
Further insight may gained in the future by applying the coincidence detection method used in our experiment to time-resolved studies on the \OO formation process from \COO with a pump-probe scheme.
Our results may trigger further experimental and theoretical investigations on the mechanism of \OO production from \COO and the optimization of the \OO production efficiency with parameters of laser pulses.

Moreover, our observation on the \OOp production from \COOpp can also provide useful hints and new concepts for the studies on planetary atmospheres. In planetary atmospheres, the existing \COO can be ionized via absorbing high energy photons or colliding with high energy electrons, which then might lead to the \OO formation like in our observation. For example, the abundance of \OO measured on Mars by the Herschel spacecraft \cite{hartogh2010herschel} and the Curiosity rover \cite{mahaffy2013abundance} may contain the contribution of \OO formation in the \COOpp layer which is predicted to be existing in the atmosphere of Mars \cite{Witasse2002}.

This research was financed by the Austrian Science Fund (FWF) under grants P25615-N27, P28475-N27, P21463-N22, P27491-N27, P25827 and SFB-F49 NEXTlite, and by a starting grant from the European Research Council (ERC project CyFi).
We also acknowledge the Vienna Scientific Cluster (VSC) for providing computational resources and the COST actions CM1204 (XLIC) and CM1405 (MOLIM).

\end{document}